\documentclass[prd,aps,amsmath,amssymb,letter]{revtex4}
\usepackage{graphicx}
\usepackage{epsfig,rotate,latexsym}
\begin{document}
\title{Super-horizon fluctuations and acoustic oscillations 
in relativistic heavy-ion collisions}
\author{Ananta P. Mishra}
\email {apmishra@iopb.res.in}
\author{Ranjita K. Mohapatra}
\email {ranjita@iopb.res.in}
\author{P. S. Saumia}
\email {saumia@iopb.res.in}
\author{Ajit M. Srivastava}
\email{ajit@iopb.res.in}
\affiliation{Institute of Physics, Sachivalaya Marg, 
Bhubaneswar 751005, India}
%
%

\begin{abstract}
 We focus on the initial state spatial anisotropies, originating at the 
thermalization stage, for central collisions in relativistic heavy-ion 
collisions. We propose that a plot of the root mean square values of the 
flow coefficients $\sqrt{\overline {v_n^2}} ~~\equiv v_n^{rms}$, calculated
in a laboratory fixed coordinate system, for a large 
range of $n$ from 1 to about 30, can give non-trivial information about
the initial stages of the system and its evolution. We also argue that 
for all wavelengths $\lambda$ of the anisotropy (at the surface of the 
plasma region) much larger than the acoustic horizon size $H_s^{fr}$ at the 
freezeout stage, the resulting values of $v_n^{rms}$ should be 
suppressed by a factor of order $2H_s^{fr}/\lambda$. For non-central 
collisions, these arguments naturally imply a certain amount of 
suppression of the elliptic flow. Further, by assuming that initial flow 
velocities are negligible at thermalization stage, we discuss the
possibility that the resulting flow could show imprints of coherent 
oscillations in the plot of $v_n^{rms}$ for sub-horizon modes. For 
gold-gold collision at 200 GeV/A center of mass energy, these features 
are expected to occur for n $\ge$ 5, with $n < 4$ modes showing 
suppression due to being superhorizon. This has strong similarities with 
the physics of the anisotropies of the cosmic microwave background 
radiation (CMBR) resulting from inflationary density fluctuations in the 
universe (despite important differences such as the absence of gravity effects
for the heavy-ion case). It seems possible that the statistical fluctuations 
due to finite multiplicity may not be able to mask such features in the 
flow data, or, at least a non-trivial overall shape of the plot of 
$v_n^{rms}$ may be inferred. In that case, the successes of analysis 
of CMBR anisotropy power spectrum to get cosmological parameters can be 
applied for relativistic heavy-ion collisions to learn about various 
relevant parameters at the early stages of the evolving system. 
\end{abstract}
\maketitle

\section{Introduction}

 In the experimental search of the deconfined phase of QCD, namely the 
quark-gluon plasma (QGP), one of the most important results has been
the observation of the elliptic flow \cite{flow0,flow1}. This has given 
strong evidence of very early thermalization and of the collective 
behavior of the partonic matter produced in RHICE \cite{flowex}.
(We use RHICE to denote general class of relativistic heavy-ion collision 
experiments, to distinguish from the Relativistic Heavy-Ion Collider, 
RHIC, at Brookhaven). Much work has been done to extract 
physical information about the system from the behavior of the elliptic 
flow, e.g., equation of state, thermalization time, freezeout time etc.
 
  Elliptic flow results from the spatial anisotropy of the thermalized
region at the initial stage in a given event with nonzero impact parameter. 
Anisotropic pressure gradients then lead to anisotropic fluid velocity 
which results in anisotropic momentum distribution of particle momenta. 
Elliptic flow measures the second Fourier coefficient of the angular 
distribution of the particle momenta in the transverse plane. It has also 
been noticed \cite{hijing,cntrl,v2cntrl} that even in central collisions, 
due to initial state fluctuations, one can get non-zero anisotropies in
particle distribution (and hence in final particle momenta) in a given event, 
though these will be typically much smaller in comparison to the non-central
collisions. These will average out to zero when large number of central
events are considered. Fluctuations in the elliptic flow resulting from
these initial state fluctuations, as well as the resulting modifications in
the eccentricity of the initial region have been investigated in the
literature \cite{eccflct,v2hyd}. In this paper, we present a different
approach to analyze the flow anisotropies for central events, resulting 
from these initial state fluctuations.

  It is sometimes mentioned in popular terms that attempts to learn
about early stages of phases of matter in RHICE from the observations 
of hadrons is in some sense similar to the attempts to understand
the early stages of the universe from the observations of the cosmic
microwave background radiation (CMBR). The surface of last scattering
for CMBR is then similar to the freezeout surface in RHICE. We will
argue below that this correspondence is in fact much deeper. There
are strong similarities in the nature of density fluctuations 
in the two cases (with the obvious difference of the absence of
gravity effects for RHICE). Following the successes of the analysis of
the CMBR anisotropy power spectrum, we argue below that, for central 
events in RHICE, a plot of the root mean square values of the flow 
coefficients $\sqrt{\overline {v_n^2}} ~~\equiv v_n^{rms}$, calculated
in a laboratory fixed coordinate system, for a large 
range of $n$ from 1 to about 30, can give non-trivial information about
the initial stages of the system and its evolution. In addition, we
recall that one of the most important aspects of
the density fluctuations in the universe is the coherence effect
which eventually results in the remarkable acoustic peaks in the
power spectrum of CMBR anisotropies. The source of this lies in
the inflationary origin of the density fluctuations leading to production 
of super-horizon density fluctuations. Some of these eventually re-enter 
the horizon around the decoupling stage and leave these imprints 
on CMBR anisotropies \cite{cmbr}. We will argue below that quite similarly,
super-horizon fluctuations are present in RHICE as well. However, here
they will result in non-zero flow coefficients as spatial anisotropies
lead to anisotropic pressure gradients. The anisotropies
in the momentum distributions of the particles, especially at large
orders of the Fourier coefficients, will capture information about the 
nature  of initial spatial anisotropies, their evolution, and freezeout.

  The paper is organized as follows. In section II we discuss the nature
of the anisotropies present initially at the stage of equilibration. 
Section III discusses the basic physics of our model 
exploring the correspondence with CMBR physics. Section IV incorporates
the presence of a finite acoustic horizon in RHICE and discusses the
expected suppression of the values of $v_n^{rms}$.
Section V discusses expected features in the plot of $v_n^{rms}$ based
on the physics of our model. Section VI presents numerical results and 
conclusions are given in section VII.

\section{Initial fluctuations in central collisions}

  As mentioned in the Introduction, main focus of our analysis is
on central events, the considerations can be trivially
extended to the non-central case. Just as for the elliptic flow,
here also we only consider transverse fluctuations by assuming
Bjorken scaling for the longitudinal expansion. This is a reasonable
approximation if the  freezeout, when momentum anisotropies of the particles 
are frozen out, does not happen too late. For a given central event, azimuthal
distribution of particles and energy density are in general anisotropic due
to fluctuations of nucleon coordinates as well as due to localized nature 
of parton production during initial nucleon collisions. As an example, for 
Au-Au collision at 200 GeV/A center of mass energy, we show the contour plot 
of initial transverse energy density in Fig.1. This is obtained using HIJING 
\cite{hijingp}. For parton positions we use random 
locations inside the volume of the parent nucleon. For partons which 
are produced by the string systems we position them randomly along the line 
joining the two nucleons corresponding to the relevant string. The 
transverse energy density at a given transverse position ${\vec x}$, at
proper time $\tau = \tau_{eq}$, is taken as \cite{hijing},

\begin{equation}
\epsilon_{tr}({\vec x},\tau_{eq}) = {1 \over \Delta A} 
\sum_i E_{tr}^i ~F(\tau_{eq},p_{tr})~
\delta^2({\vec x} - {\vec x}^i_0 - {\vec v}^i \tau_{eq}) ~\Delta(y^i)
\end{equation}

where ${\vec x}^i_0$ denotes the initial transverse coordinates of the
$i_{th}$ parton (determined using the coordinates of the parent nucleon in
HIJING as discussed above), $E^i_{tr}$ is its transverse energy, $p_{tr}$
the transverse momentum, and ${\vec v^i}$ is its transverse velocity. 
For the rapidity window we take $\Delta(y^i) = 1$ centered at $y = 0$, 
\cite{hijing}. The sum over $i$ includes all partons in a small 
transverse area element $\Delta A (\simeq 0.5$ fm$^2$) at position 
${\vec x}$. Following ref.\cite{hijing,fact} we have included a 
factor $F(\tau_{eq},p_{tr}) \equiv 1/(1 + 1/(p_{tr} \tau_{eq})^2)$ to 
account for the probability of formation of partons with zero rapidity.
Fig.1 shows the contour plot of the
energy density at $\tau_{eq}$ = 1 fm. We assume that by this time the 
produced partons thermalize and hydrodynamic description becomes 
applicable for subsequent times, (with the energy density decreasing,
for $\tau > \tau_{eq}$, due to longitudinal as well as transverse
hydrodynamical expansions). We will present results with the transverse 
energy given as above in Eq.(1). We have also checked the effects of
the presence of a smooth background, such as the one representing
soft beam-jet component in ref.\cite{hijing}. As in ref.\cite{hijing},
we model it as $\epsilon_{soft} \sim \rho_c (1 - (r_{tr}/R)^2)^{1/2}$,
where $r_{tr}$ is the transverse radial coordinate, and $R$ is taken
as the nucleus size. The central density for this soft component,
$\rho_c$, at $\tau_{eq}$, is varied from 0 up to about 3 GeV/fm$^2$ 
\cite{hijing}. As expected, the presence of this soft component reduces 
the overall magnitude of the fluctuations, and hence the initial 
anisotropies. However, the qualitative nature of our results do 
not change with this, as we will discuss later in Sect.IV. 

\begin{figure}
\vskip -1.0in
\epsfig{file=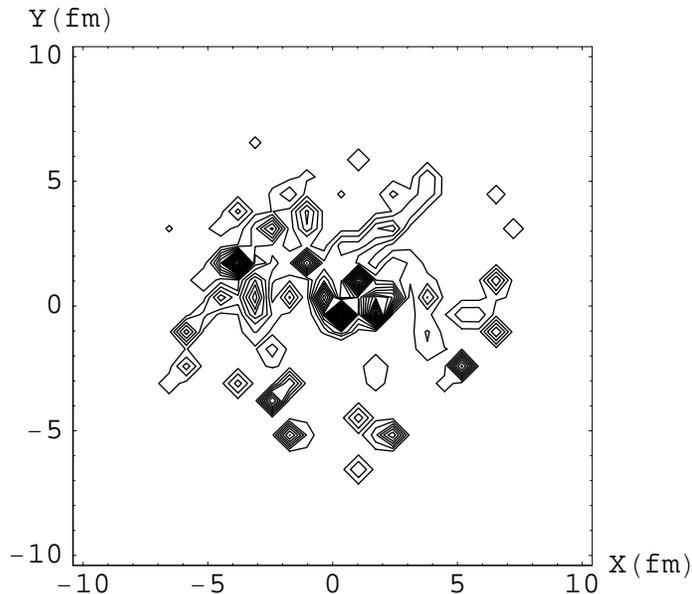,height=175mm}
\vskip -2.5in
\caption{Contour plot of the initial transverse energy density 
distribution from HIJING for a given $Au-Au$ collision central 
event at 200 GeV/A center of mass energy}
\label{fig:fig1}
\end{figure}

Azimuthal anisotropy of produced partons is manifest in Fig.1.
It is thus reasonable to expect that the equilibrated matter resulting
from this parton distribution will also have azimuthal anisotropies
(as well as radial fluctuations) of similar level. The process of 
equilibration will lead to some level of smoothening. However, flow
measurements have given very interesting result that thermalization is 
expected to happen quickly, within proper time  $\tau_{eq} \le$ 1 fm for 
RHIC energies. Hence, no 
homogenization can be expected to occur beyond length scales larger than 
this, leading to presence of non-uniformities at the initial stage. 

\section{Basic physics of the model}

 From the discussion in the last section we conclude that inhomogeneities, 
especially anisotropies with wavelengths larger than
the thermalization scale should be necessarily present at the thermalization
stage when the hydrodynamic description is expected to become applicable.
This brings us to the most important correspondence between the universe
and the RHICE. It is the presence of fluctuations with superhorizon
wavelengths. In the
universe, density fluctuations with wavelengths of superhorizon scale have 
their origin in the inflationary period. Quantum fluctuations of sub-horizon
scale are stretched out to superhorizon scales during the inflationary
period. During subsequent evolution, after the end of the inflation,
fluctuations of sequentially increasing wavelengths keep entering the 
horizon. The largest ones to enter the horizon, and grow, at the stage of 
decoupling of matter and radiation lead to the first peak in CMBR anisotropy 
power spectrum. 

   Fig.1 shows that superhorizon fluctuations should be present in RHICE at 
the initial equilibration stage itself. The reason these are superhorizon is 
that the causal horizon is given by c$\tau$.  More appropriately, one should
be using the sound horizon, $ H_s \sim c_s \tau$ where $c_s$ is the sound 
speed, as we are interested in the flow arising from pressure
gradients.  At the stage of equilibration, c$\tau_{eq}$ is at most 1 fm, 
with corresponding acoustic horizon $H_s^{eq}$ being even smaller. Thus every  
fluctuation of wavelength larger than $H_s^{eq}$ is superhorizon. With
the nucleon size being about 1.6 fm, the equilibrated matter will
necessarily have density inhomogeneities with superhorizon wavelengths. As 
the system evolves beyond $\tau_{eq}$, fluctuations of larger wavelengths 
enter the horizon. We will argue in section IV that the (r.m.s. values of) 
the flow coefficients corresponding to fluctuations which remain superhorizon
at the freezeout stage will be suppressed.

It is important to note that density fluctuations will also be present in 
the longitudinal direction. However, partons being created near $z = 0$ 
($z$ being the longitudinal coordinate), and subsequently expanding out, 
one cannot argue that these longitudinal fluctuations are out of causal 
contact (except possibly, in the color glass condensate models \cite{cgc},
as we discuss later). In contrast, the transverse 
fluctuations are superhorizon because of simultaneous collisions of
different (transverse) parts of a system with large transverse dimensions
compared to the causal horizon at the thermalization stage (the system being 
longitudinally Lorentz contracted nuclei, or at the smallest level, colliding 
nucleons or clusters of nucleons). In this way, the fluctuations 
leading to the elliptic flow are necessarily superhorizon. As we will see 
later, they remain superhorizon even at the freezeout stage, and hence the 
resulting elliptic flow is necessarily suppressed compared to maximum possible
value. This difference between the longitudinal and transverse fluctuations
may be important in the context of our model (apart from other differences 
due to different expansion dynamics in the two directions), as we will 
discuss later. 

 To estimate spatial anisotropies for the system as in Fig.1 we will
use the following procedure. As mentioned, we assume that the
hydrodynamic description becomes applicable by $\tau =\tau_{eq}$,
which we take to be 1 fm. We calculate the anisotropies in the 
fluctuations in the spatial extent $R(\phi)$ at this stage, where $R(\phi)$
represents the energy density weighted average of the transverse
radial coordinate in the angular bin at azimuthal coordinate $\phi$. 
We divide the region in 50 - 100 bins of azimuthal angle $\phi$, and 
calculate the Fourier coefficients of the anisotropies in ${\delta R}/R 
\equiv  (R(\phi) - {\bar R})/{\bar R}$ where $\bar R$ is the angular 
average of $R(\phi)$. Note that in this way we are representing
all fluctuations essentially in terms of fluctuations in the boundary of 
the initial region. Clearly there are density fluctuations in the interior
region as well. However, in view of thermalization processes operative
within the plasma region, as far as the development of flow anisotropies
is concerned, presumably the representation by fluctuating boundary will
capture the essential physics. A more careful analysis should include
the details of fluctuations in the interior regions and their effects
on the evolution of the flow. We will use $F_n$ to denote Fourier
coefficients for these spatial anisotropies, and use the conventional
notation $v_n$ to denote $n_{th}$ Fourier coefficient of the resulting
momentum anisotropy in ${\delta p}/p$. Here $\delta p$ represents 
fluctuation in the momentum $p$ of the final particles from the average 
momentum, in a given azimuthal angle bin. 

 An important difference between the conventional discussions of the 
elliptic flow and our analysis is that, here
one does not try to determine any special reaction plane on event-by-event
basis. A fixed coordinate system is used for calculating azimuthal
anisotropies. This is why, as we will see later, averages of $F_n$s
(and hence of $v_n$s) will vanish when large number of events are included
in the analysis. However, the root mean square values of $F_n$s, and hence of
$v_n$s, will be non-zero in general and will contain non-trivial 
information. In fact, it is the same as the standard deviation for the 
distribution of $F_n$s since the average value of $F_n$s is zero. This is 
what is exactly done for the CMBR case also \cite{cmbr}. This is why even when
temperature fluctuations are very tiny for CMBR (1 part in 10$^5$), one is
still able to resolve the acoustic peaks. Note, we use $v_2$ with the 
present definitions to denote the  elliptic flow even though we do not adopt 
the conventional usage of the eccentricity for defining the corresponding 
spatial anisotropy. Clearly the procedure described above gives a crude 
estimate of the spatial anisotropy of the initial plasma region. We have 
taken  other different measures of the spatial anisotropy and, as we will 
discuss later, it does not affect our results much.

\subsection{Correspondence with CMBR physics: Coherence of fluctuations}

   Before we proceed any further in analyzing the nature of density
fluctuations in RHICE and its similarities with CMBR anisotropies, let
us be clear about the relevant experimentally measurable quantities.
For the case of the universe, density fluctuations at the surface of last
scattering are accessible through their imprints on the CMBR. 
Thus only for the fluctuations present at the decoupling stage,
starting from the short wavelength fluctuations which would have
undergone several oscillations until decoupling, including the large
wavelength fluctuations which just enter the horizon at decoupling 
and start growing due to gravity, upto the superhorizon fluctuations,
CMBR anisotropies capture imprints
of all of them, and are observed today \cite{cmbr}.  For the RHICE case, 
the experimentally accessible data is particle momenta which are finally 
detected. Initial stage spatial anisotropies are accessible only as long
as they leave any imprints on the momentum distributions (as for the
elliptic flow) which survives until the freezeout stage. What one is looking
for, therefore, is the evolution of spatial anisotropies of different
wavelengths, and corresponding  buildup of momentum anisotropies
(i.e. essentially different flow coefficients), existing at the freezeout
stage.  

   The two most crucial aspects of the inflationary density fluctuations
leading to the remarkable signatures of acoustic peaks in CMBR are
coherence and acoustic oscillations. Let us consider them in turn to see
if any such features are expected for RHICE. Let us recall that coherence
of inflationary density fluctuations essentially results from the fact
that the fluctuations initially are stretched to superhorizon sizes and are 
subsequently frozen out dynamically. Thus, at the stage of re-entering the 
horizon, when these fluctuations start growing due to gravity, and 
subsequently start oscillating due to radiation pressure, the fluctuations 
start with zero velocity. For an oscillating fluctuation, this will mean 
that only $cos\omega t$ term survives. As all the fluctuations,
entering the horizon at a given stage, have same wavelength 
(by definition) they all are phase locked, coming to zero amplitude 
simultaneously. In summary, the crucial requirement for coherence of 
fluctuations is that they are essentially frozen out until they re-enter 
the horizon \cite{cmbr}.

  This should be reasonably true for RHICE, especially as we are
considering transverse fluctuations. Transverse velocity to begin with
is expected to be zero. Though, note that with initial state fluctuations 
due to fluctuations in nucleon (hence partons) positions and momenta, there 
may be some residual transverse velocities even at the earliest stages 
\cite{hijing}. However, due to averaging, for wavelengths significantly
larger than the nucleon size, it is unlikely that the fluid will
develop any significant velocity at the thermalization stage.
For much larger wavelengths, those which enter (sound) horizon at
proper times much larger than $\tau_{eq}$, build up of the radial expansion 
will not be negligible. However, our interest is in the presence of any 
oscillatory modes. For such oscillatory time dependence even for
such large wavelength modes, there is no reason to expect the presence
of $sin\omega t$ term at the stage when the fluctuation is just entering
the sound horizon.   

\subsection{Acoustic oscillations}

  Let us now address the possibility of the oscillatory behavior for the
fluctuations. In the universe, attractive forces of gravity and counter
balancing forces from radiation pressure (with the coupling of baryons to 
the radiation) lead to acoustic oscillations \cite{cmbr}. For RHICE, there 
is no gravity, but there is a non-zero pressure present in the system.
First let us just follow the conventional
analysis as in the case of elliptic flow. We know that, as the spatial
anisotropy decreases in time by the buildup of momentum anisotropy (starting
from isotropic momentum distribution), it eventually crosses zero and becomes
negative. This forces momentum anisotropy to saturate first (when spatial
anisotropy becomes zero), and then start decreasing. In principle, one
could imagine momentum anisotropy to decrease to zero, becoming negative
eventually. The whole cycle could then be repeated, resulting in an
oscillatory behavior for the spatial anisotropy as well as for the momentum
anisotropy, possibly with decreasing amplitude. 

   Unfortunately, the situation is not that favorable. For the elliptic
flow, in hydrodynamic simulations one does see saturation of the flow, 
and possibly turn over part \cite{vtr}, but the momentum anisotropy does 
not become zero, and there is never any indication of an oscillatory
behavior. Important thing to note here is that this primarily happens
because of the build up of the strong radial flow by the time elliptic
flow saturates, and subsequently freezes out. This can be seen from the
evolution of transverse velocity in hydrodynamics simulations \cite{vtr}.
One then would like to know whether the same fate is necessarily true for 
fluctuations of much smaller wavelength as well.
At this stage it is important to be clear about the relevant
freezeout time $\tau_{fr}$  for the flow. If radial expansion becomes very 
strong then pressure gradients, or any interface tension effects (as we will
discuss below) may not be able to significantly affect fluid flow
anisotropies. Flow anisotropy would then essentially freezeout 
even if chemical or thermal freezeout may not have occurred yet. For 
elliptic flow this is usually accounted for by referring to the flow 
build up time scale of order $ R/c_s$ where $R$ represents the initial 
average transverse extent of the region \cite{flow0}. This is the time 
scale when transverse expansion is expected to become strong.  For central 
collisions, we will take $R = {\bar R}(\tau_{eq}) \equiv {\bar R}$ which 
was earlier defined as the energy density weighted average of the transverse 
radial coordinate at $\tau = \tau_{eq}$. For $Au-Au$  collision at 200 GeV,
the value of ${\bar R}$ is obtained to be about 3 fm from HIJING. This 
gives us $\tau_{fr} = {\bar R}/c_s + \tau_{eq} \simeq 6$ fm. (We use
velocity of sound $c_s = 1/\sqrt{3}$. Later we will discuss the effects
of changing the value of $c_s$.) Note that this value of $\tau_{fr}$
is much smaller than the value of the thermal freezeout time at 
these energies which is expected to be of order 12 fm.  Certainly flow 
anisotropies have to freeze by the stage of thermal freezeout. We have 
also used the value of $\tau_{fr}$ to be 12 fm and it does not change our 
results, as we will discuss later. We will use the freezeout stage as given 
by $\tau_{fr} - \tau_{eq} = {\bar R}/c_s$. The size of the acoustic horizon 
at $\tau = \tau_{fr}$ is then simply $H^{fr}_s = c_s (\tau_{fr} - 
\tau_{eq}) = {\bar R} \simeq 3$ fm for $Au-Au$ collision at 200 GeV. 
We take the sound horizon to be proportional to the proper time elapsed 
since the stage of equilibration given by $\tau_{eq}$, because this is the 
stage when hydrodynamics becomes applicable. Before $\tau_{eq}$, individual 
particles can interact with each other, so in principle one could define
a non-zero causal horizon size for individual particles. However, there is 
no sense in which one can talk about the interaction of different collective 
modes before $\tau_{eq}$. Hence it seems reasonable that the acoustic 
horizon, as appropriate for these collective modes, is defined in terms
of time elapsed from $\tau_{eq}$. In the present case with $\tau_{eq}$ 
being very small, of order 1 fm, our results remain almost unchanged even 
if we take $H^{fr}_s = c_s \tau_{fr}$. Important point to note is that this 
time $\tau_{fr}$ remains fixed for a given collision, irrespective of the
wavelength $\lambda$ of the fluctuation considered. For small wavelengths,
if the time scale for the build up of momentum anisotropy is much smaller 
than $\tau_{fr}$, then oscillations may be possible before the flow freezes 
out at $\tau_{fr}$. Consider spatial anisotropy with a wavelength which is 
much shorter than $H^{fr}_s$ at the freezeout stage, say $\lambda$ being of 
order 2 fm, see Fig.2. One will expect that due to unequal initial pressure 
gradients in the two directions $\phi_1$ and $\phi_2$, momentum anisotropy 
would have built up in these two directions in relatively short time. 
Most importantly, we expect that spatial anisotropy should
reverse sign in time of order $\tau_{flip} \simeq \lambda /(2c_s) \simeq 2$ 
fm. The momentum anisotropy should then reach saturation before this time, 
and start decreasing by  $\tau \simeq \tau_{flip}$. Due to short time scale 
of evolution here, radial expansion may still not be most dominant and there 
may be possibility of momentum anisotropy changing sign, leading to some 
sort of oscillatory behavior.

\begin{figure}
\vskip -2.0in
\epsfig{file=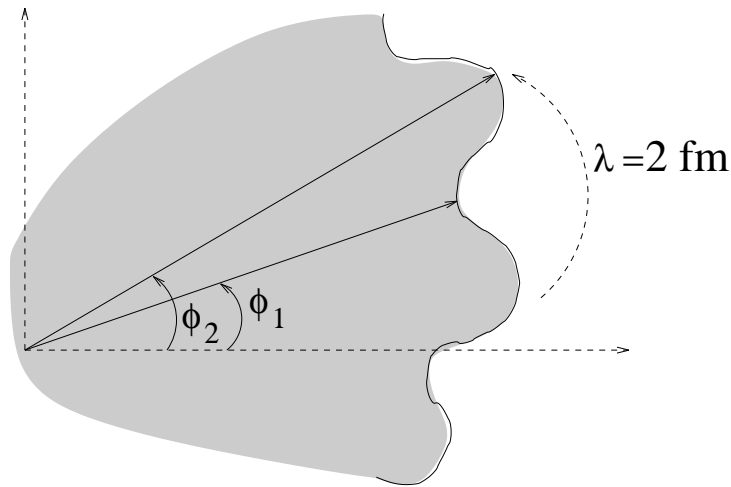,height=175mm}
\vskip -2.2in
\caption{Schematic diagram of a part of the region showing spatial 
anisotropies of small wavelength}
\label{fig:fig2}
\end{figure}

   The arguments presented above are very qualitative, but they allow 
a possibility of oscillatory behavior for the spatial anisotropy (i.e. 
fluctuations) of short wavelengths, and hence possibly for the flow anisotropy. 
An entirely different line of argument supporting oscillatory behavior can 
be given as follows. As the QGP region is surrounded by the confining 
vacuum (or the hadronic phase), one may expect some sort of 
interface, with non-zero surface tension, to be 
present at the boundary of the QGP region \cite{digal}. This is necessarily 
true if the confinement-deconfinement transition is of first order.
Once we assume an interface present at the boundary of the QGP region,
its surface tension will induce oscillatory evolution given initial
fluctuations present (as shown in Figs.1,2). For a relativistic domain
wall such fluctuations propagate with speed of light. However, here
one will be dealing with an interface bounding a dense plasma of quarks
and gluons. In such a case it will be reasonable to expect that tiny
perturbations on the interface will evolve with speed of sound (appropriate
for the QGP region enclosed). The interface will eventually disappear
after the quark-hadron phase transition stage. We mention here that
the contributions of any such interface could also be present (though,
possibly small, relatively) for the
conventional calculations of the elliptic flow for non-central collisions.
Also, effects of surface tension will be more prominent for lower
energy collisions, and the considerations of our model can be applied
for such low energy heavy-ion collision experiments.
 
\section{Acoustic horizon and development of flow}

  One important aspect of these fluctuations is that of {\it horizon
entering}. In the case of the universe every fluctuation leads to CMBR
anisotropies even when it is superhorizon. This is because these are density
fluctuations and local thermal equilibrium directly leads to associated
CMBR temperature fluctuations. The importance of horizon entering there is
for the growth of fluctuations due to gravity. This leads to increase in the
amplitude of density fluctuations, with subsequent oscillatory evolution,
leaving the imprints of these important features in terms of acoustic
peaks \cite{cmbr}. For RHICE, there is a similar (though not the same, due 
to absence of gravity here) importance of horizon entering. Azimuthal spatial 
anisotropies are not directly detected. They are detected only when they
are transferred to the momentum anisotropies of particles. Consider first
an anisotropy of wavelength $\lambda$ much larger than the acoustic horizon 
at freezeout $H^{fr}_s$. The evolution of fluid velocity (particle momenta) 
depends on pressure gradients.
However, pressure gradients at $\tau = \tau_{fr}$  should only be calculated
within regions of size $H^{fr}_s$ as outside regions could not have affected
the region under consideration.  Consider, for
simplicity, fluid expanding radially with initial transverse velocity $v$
being zero. Then keeping only terms which are of first order in $v$
(and neglecting $\partial P /\partial \tau$ term) the Euler's 
equation gives \cite{euler},

\begin{equation}
{\partial v \over \partial \tau} = {-1 \over (\rho + P)} 
{\partial P \over \partial r}
\end{equation}

where $\rho$ is the energy density, $P$ is the pressure, and $r$ is the
transverse radial coordinate. This equation illustrates the expected build 
up of the flow velocity being proportional to the pressure gradient. 
However, for a spatial anisotropy with $\lambda >> H_s^{fr}$ the angular 
variation of the pressure gradient should only be calculated
using information within a region of size $H^{fr}_s$. We will argue below
that we can approximate the angular variation of the relevant pressure 
gradient, operative  within length scale of $H^{fr}_s$, to be a fraction 
of order $H^{fr}_s/(\lambda/2)$ of its  value arising from the full magnitude 
of the spatial anisotropy. 

 Presence of such a suppression factor is most naturally seen for the case 
when the build up of the anisotropies in the flow from spatial anisotropies 
is dominated by the surface tension of the interface bounding the QGP region.
For an anisotropy of wavelength $\lambda$, the maximum and minimum
of spatial extent are separated by distance $\lambda/2$ at the surface (see, 
Fig.2). After half oscillation we expect the reversal of the anisotropy,
i.e. maximum will become minimum and minimum will become maximum. However,
if the acoustic horizon $H^{fr}_s = c_s (\tau_{fr} - \tau_{eq}) 
<< \lambda/2$, then this full reversal is not possible. Starting from the 
position of the initial maximum amplitude, the relevant amplitude for 
oscillation is only a factor of order $H_s/(\lambda/2)$ of the full 
amplitude. Thus, we conclude that the resulting anisotropy of the flow for 
this particular mode will be suppressed by a factor $f$ of order 
$2H^{fr}_s/\lambda$.
 
Let us now consider the situation when flow anisotropies result from the 
pressure gradients, as in the conventional calculations of the elliptic 
flow for non-central events. There, the difference in
the pressure gradients in the two directions (X and Y) arises from
the difference in the values of the semi-minor axis a and semi-major axis b 
of the elliptical transverse overlap region. A given central pressure $P_c$
then gives different pressure gradients in the X and Y directions. However,
it should be obvious that for times at which the acoustic horizon is
much smaller than a, or b, the value of $P_c$ should be irrelevant. There 
is no way that the surface of QGP would get any information about $P_c$ 
for $H_s = c_s \tau << a,b$. Only relevant pressure gradients should be 
calculated using interior pressure value at distances of order $c_s \tau$ 
starting from the surface in the X and the Y directions. For example, for
our case of central events, let us parametrize the radial profile of the 
pressure, along a particular direction $\phi$, as 

\begin{equation}
P(r) = P_c exp[{-r^2 \over 2 \sigma_{\phi}^2}]
\end{equation}

with $\sigma_{\phi}$ varying with azimuthal angle giving the spatial 
anisotropy. For calculating the pressure gradient at small time $\tau$
(measured from $\tau_{eq}$), at $r = R_0$ at the surface region, we can 
use the following estimate.

\begin{equation}
{\partial P \over \partial r} \simeq {P(R_0 - c_s \tau) - P(R_0)
\over (-c_s \tau)} \simeq -P_c exp[{-R_0^2 \over 2 \sigma_{\phi}^2}] 
\left ({R_0 \over \sigma_{\phi}^2} + {c_s \tau \over 2 \sigma_{\phi}^2}
({R_0^2 \over \sigma_{\phi}^2} -1) \right)
\end{equation}

 This example shows that $-\partial P/\partial r$, at the surface for 
$R_0 > \sigma_{\phi}$, starts from a small value at $\tau = 0$ (again 
$\tau$ measured from $\tau_{eq}$), and increases with the sound  horizon 
as $c_s \tau$. Though, the growth of velocity also requires 
the value of  $\rho + P$ (whose average value also increases with 
$\tau$), which in some sense provides the inertia of the fluid element. 
However, we can focus on a particular fluid element near the surface 
and estimate its acceleration using above pressure gradient. Thus, 
though we are calculating the average pressure gradient using an 
increasing region with $\tau$ (as $c_s \tau$), we do not consider the 
average velocity of this entire region. We are always focusing on the 
region near $r \simeq R_0$ near the surface. The fact that the magnitude of 
the average pressure gradient increases in time suggests that the pressure
gradient operative at $r \simeq R_0$ should also increase in time.
Note that Eq.(3) is only meant to give the initial profile of the
pressure and it would be incorrect to use it to calculate $\partial P
/\partial r$ (say, at $R_0$) at a later stage, by which the interior higher 
pressure will be expected to affect the fluid flow at the surface.
(For $\tau = 0$, Eq.(4) gives the same result for $\partial P /\partial r$
as directly obtained  from Eq.(3).)
Also note that the particular dependence on $\tau$ in Eq.(4) is due to 
our specific choice of pressure profile in Eq.(3). For example, a linear 
$r$ dependence for the pressure in Eq.(3) would give constant pressure 
gradient in time for Eq.(4) by the above estimate. However, the pressure 
profile of the sort in Eq.(3) is more representative of the expected 
profile than a linear dependence. (Though, even here, for interior regions 
with $R_0 < \sigma_{\phi}$, one will conclude 
that $-\partial P /\partial r $ decreases in time.) We mention that the 
physical considerations discussed above may be relevant for the early rise 
of the pressure gradients seen in the simulations in ref. \cite{pgrad}. 

  The length scale relevant for considering the azimuthal variation of the 
flow is the wavelength $\lambda$ of the fluctuation under consideration. 
Again, for a spatial anisotropy of wavelength $\lambda$ near the surface of 
the QGP region, we would like to know flow buildup at two regions separated 
by $\lambda/2$. These two regions will have access to a common value of 
central pressure after a time $\tau_{cmn}$ with $\tau_{cmn} - \tau_{eq} 
\simeq \lambda/(2 c_s)$. This implies that as far as azimuthal
anisotropy of the pressure gradient, and the resulting flow, are concerned,
there should be no suppression in the anisotropy after time $\tau_{cmn}$
arising from the considerations of horizon entering. In some sense, this
fluctuation would be said to have "entered the horizon" after $\tau_{cmn}$.
For $\tau_{cmn} = \tau_{fr}$ this means  that the largest wavelength mode 
which will be unsuppressed,  will have $\lambda_{max} \simeq  2 c_s 
(\tau_{fr}-\tau_{eq})$.

The above discussion leads us to the conclusion that for 
superhorizon fluctuations (at freezeout stage), full momentum
anisotropy $(v_n)_{max}$, as expected from hydrodynamics, will not be 
developed, and only a fraction $f$ of the maximum possible anisotropy 
will develop. In terms of the corresponding flow coefficients, 
for modes with $\lambda/2 \ge H^{fr}_s$, we expect,

\begin{equation}
(v_n)_{observed} = {2H^{fr}_s \over \lambda} (v_n)_{max}
\end{equation}

where $\lambda \sim 2\pi {\bar R}^{fr}/n, ~~~(n \ge 1)$, is the measure of the 
wavelength of the anisotropy corresponding to the $n_{th}$ Fourier coefficient.
Here $\bar R^{fr}$ represents the transverse radius at the stage $\tau_{fr}$.
Using the rough estimate of the rate of change of the transverse velocity
to be about 0.1 fm$^{-1}$ at the early stages at these energies \cite{vtr}, 
we can estimate $\bar R^{fr} \simeq {\bar R} + 0.05 (\tau_{fr} - 
\tau_{eq})^2 = {\bar R} (1 + 0.05 {\bar R}/c_s^2)$. Here ${\bar R} \equiv 
{\bar R}(\tau_{eq}) = c_s (\tau_{fr} - \tau_{eq})$, as discussed above. 
The largest wavelength $\lambda_{max}$ of spatial anisotropy which will 
have chance  to develop to its maximum hydrodynamic value is, therefore, 
$\lambda_{max} \simeq 2 H^{fr}_s = 2 c_s (\tau_{fr} - \tau_{eq})
= 2 {\bar R}(\tau_{eq})$. This gives us the corresponding minimum value 
$n_{min}$ of $n$ below which flow coefficients should show suppression
due to being superhorizon,

\begin{equation}
n_{min} = \pi (1 + {0.05 {\bar R}(\tau_{eq}) \over c_s^2} )
\end{equation}

\section{Expected features in plots of $v_n^{rms}$}

 First we would like to emphasize the important lessons from
CMBR analysis which we are proposing to use for RHICE.
As we are considering central events, associated anisotropies will be 
small. Thus we will need large amount of statistics to be able to see any 
non-trivial features, especially the possibility of oscillations
etc. An important thing to realize is that we cannot take averages
of the Fourier coefficients, as these will simply be zero when large
number of events are included. Note that our analysis is being done
in a laboratory fixed frame. For elliptic flow, one defines
an event plane and a special direction given by the impact parameter
and then calculates average spatial anisotropy. There is no such direction 
for a central event. The correct measure of the spatial anisotropy here is
to take the root mean square values of the Fourier coefficients taken for the 
whole ensemble of events. As the average value of $F_n$s is zero, it is the 
same as the standard deviation for the distribution of $F_n$s. This is what is 
exactly done for the CMBR case \cite{cmbr}. This is why even when
temperature fluctuations are very tiny for CMBR (1 part in 10$^5$), one is
still able to resolve the acoustic peaks. One important difference which
is in the favor of RHICE is the fact that for the CMBR case, for each $l$ mode
of the spherical harmonic, there are only $2l+1$ independent measurements 
available, as there is only one CMBR sky to observe. In particular for small 
$l$ values this leads to the accuracy limited by the so called cosmic 
variance \cite{cmbr}. In contrast, for RHICE, each  nucleus-nucleus 
collision (with same parameters like collision energy, centrality etc.) 
provides a new sample event (in some sense like another universe). The 
accuracy, for any Fourier mode, is only limited here by the number of 
events one includes in the analysis. Therefore it should be possible to
resolve any signal present in these events as discussed in this paper.
Note, due to the absence of any special reflection symmetry here (which was
present in the elliptic flow case) there is no reason to expect
that only even flow coefficients will be present. In our case all flow 
coefficients give non-zero contributions to $v_n^{rms}$, and the $``sin''$ 
terms give same values as the $``cos''$ terms. In the plots of $v_n^{rms}$
in the next section, we show the sum of these two contributions, i.e.
square root of the sum of the squares of the $``sin"$ term and the 
$``cos"$ term.

 We have argued in the last section that for fluctuations with $n < 
n_{min}$ (Eq.(6)) we expect suppression of values of $v_n^{rms}$.
For the $Au-Au$ collision at 200 GeV, with ${\bar R}(\tau_{eq}) \simeq$
3 fm, we get $n_{min} \simeq 4.5 \simeq$ 5. Note in particular that this 
implies that the mode with $n = 2$, which corresponds to the elliptic 
flow, will be expected to be suppressed by a factor of order $f \simeq 1/2$. 
This suppression is roughly of same order as the suppression factor 
for the elliptic flow discussed in the literature \cite{v2supr,v2hyd}.
Note, however, that in ref.\cite{v2supr,v2hyd}, the suppression of the
elliptic flow is related to the non-zero value of the Knudsen number 
arising from incomplete thermalization. One will then expect that the final 
suppression should be a combination of both of these factors. Note, also, 
that if velocity of sound $c_s$ becomes smaller then, taking flow freezeout to 
occur at fixed time, the acoustic horizon $H_s$ will be smaller. Suppression
factor in Eq.(5) will be stronger then leading to a smaller flow. We can 
also see from Eq.(6) that decreasing $c_s$ increases $n_{min}$ leading to 
larger suppression for the elliptic flow (by a factor $\simeq 2/n_{min}$). 
This is consistent with the findings in \cite{v2hyd}.
We mention here that the scaling of 
$v_2$ with $c_s (t - t_0)$ is known in literature, see ref. \cite{v2hyd}.
What we have discussed above essentially says that similar scaling, when
applied to spatial anisotropies with different wavelengths (i.e. different
Fourier modes), will lead to a multiplicative factor given by Eq.(5)
for the corresponding flow anisotropies.

  For all wavelengths smaller than $\lambda_{max}$, spatial anisotropy
should be able to develop into its full hydrodynamic value (apart from
the effects such as incomplete thermalization \cite{v2supr,v2hyd}, as
mentioned above). For the case of elliptic flow several studies have
shown that for near central events,
the  momentum  anisotropy is related to the initial spatial anisotropy by 
a proportionality constant of about 0.2 \cite{cntrlex}. The relation between 
the Fourier coefficients of the spatial anisotropy and resulting momentum 
anisotropy in our model can only be 
obtained using a full hydrodynamical simulation, 
with proper accounts of any surface tension, as well as factors such as 
horizon crossing etc. to properly account for the physics discussed here. 
In the absence of such a simulation, we make a strong assumption here that 
all Fourier coefficients for momentum anisotropy are related to the 
corresponding coefficients for spatial anisotropy by roughly the same 
proportionality factor, which we take to be 0.2 for definiteness.
As we will see, this choice gives us reasonably good agreement with the 
results for $v_2$ in the literature for (almost) central events. 
Further, for simplicity we evaluate the Fourier coefficients for
the spatial anisotropy, $F_n$s, at the initial stage $\tau = \tau_{eq}$. 
In principle one can consider relating the final flow coefficients $v_n$
(which freezeout at $\tau = \tau_{fr}$) to the $F_n$s at later stages.
For example we could estimate $F_n$s at $\tau = \tau_{fr}$, or more
appropriately, for each mode $n$ we could evaluate $F_n$ at the stage
of its horizon entering (meaning the stage when the sound horizon
equals $\lambda/2$ for that mode). Such estimates will require 
suitable modeling of transverse expansion and we hope to do that in a 
future work. For now we wish to emphasize that the ambiguity in relating
$v_n$ to $F_n$ does not affect the important qualitative aspects of
our results, as we will discuss later.

   As the fluctuation with $\lambda = \lambda_{max}$ will be expected to
achieve its maximum value at the freezeout, we expect that for all 
wavelengths $\lambda < \lambda_{max}$, if any oscillatory behavior is
present, as discussed above, then it will lead to an additional factor of 
$cos\omega \tau$ (with $\tau$ measured from $\tau_{eq}$) in the Fourier 
coefficient where $\omega = 2\pi c_s/\lambda$ 
is the frequency of oscillation for the relevant fluctuation. This will imply 
that the fluctuations which will have their maximum values at the freezeout
stage will have wavelengths given by $\lambda_N \sim \lambda_{max}/N$
where $N = 1,2,3...$ is a positive integer. Note, this assumes that
oscillation time scale for these fluctuations is governed by the sound
speed which will be expected if oscillations are dominated by the
interface tension at the QGP boundary. For the oscillations resulting 
from the reversal of flow due to anisotropic pressure gradients
(as discussed above), the oscillation time scale may be different.

\section{Numerical  Results}

   We are now ready to discuss our results. We have generated events using    
HIJING and we present sample results for Au-Au collision at 200 GeV/A
center of mass energy. Fig.3 shows plots where averages are taken over 
10000 events. The solid curve in Fig.3 shows the root mean square values 
$v_n^{rms}$ of the flow Fourier coefficients which are obtained from spatial 
$F_n$s using proportionality factor of 0.2 (as discussed above). $F_n$s are
calculated directly from events as depicted in Fig.1 with the rapidity 
constraints as discussed for Eq.(1). For comparison, we show here the plot
of $v_n^{rms}$ obtained from $F_n$s when partons are distributed in a nucleus 
size region with uniform probability (dotted plot in Fig.3). The $p_T$ 
distribution of partons for this dotted plot in Fig.3 is taken to be the 
same as obtained from HIJING. We see that the resulting values of 
$v_n^{rms}$ are significantly smaller than the values shown by the solid 
plot in Fig.3. Also, the plot of $v_n^{rms}$ vs. $n$ is flat as shown by 
the dotted curve in Fig.3, in contrast to the non-trivial shape of the
solid plot. We have also considered uniform energy distribution 
among partons, along with uniform probability distribution 
for their positions. This leads to even smaller values for $v_n^{rms}$
compared to the dotted curve in Fig.3 (about 15\% smaller), and the plot
remains flat. This suggests that statistical fluctuations in parton 
positions may remain subdominant and a genuine fluctuation in parton 
positions (as depicted in Fig.1) may be visible by the non-trivial shape of 
the solid plot in Fig.3. 

\begin{figure}
\vskip -2.5in
\epsfig{file=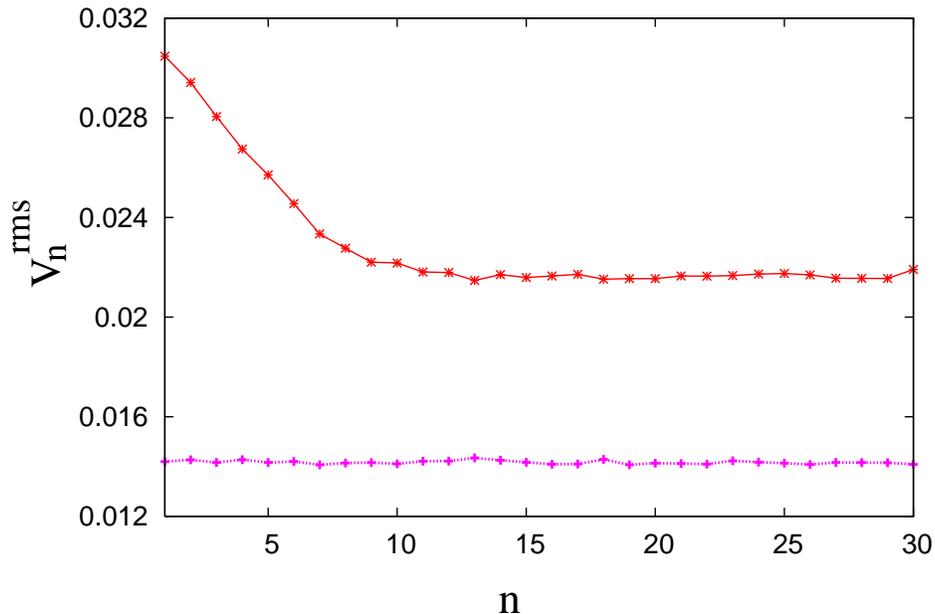,height=225mm}
\vskip -3.0in
\caption{These plots represent $v_n^{rms}$ calculated from the 
Fourier coefficients $F_n$s of the spatial anisotropy using a
proportionality factor of 0.2 as discussed in the text. Plots 
show smooth joining of the values of $v_n^{rms}$ and obtained 
using 10000 events from HIJING. Solid curve is obtained using
parton positions from HIJING as discussed for Eq.(1). 
Dotted curve is obtained with a uniform distribution of 
parton positions in a nucleus size region.}
\label{fig:fig3}
\end{figure}

We have also calculated the root mean square values of the Fourier 
coefficients of the momentum anisotropy directly using the momenta of
final particles from HIJING, using same rapidity window as for plots
in Fig.3. These are shown by the plots in Fig.4.
For the sake of consistency, we denote these also by $v_n^{rms}$.
Here averages are taken over 15000 events. The dotted plot represents 
contributions from all final particles. Note that this curve is not 
entirely flat as should be expected from an isotropic momentum 
distribution with white noise. Though fractional change for low values 
of $n$ is small compared to the solid plot in Fig.3. The plot becomes 
almost flat when only low $p_T (\le 200$ MeV) particles are included, 
as shown by the solid plot, suggesting that larger values of 
$v_n^{rms}$ at lower $n$ in the dotted plot  are due to jet correlations. 
Overall larger values of $v_n^{rms}$ for the solid plot compared 
to the dotted plot here are due to smaller number of particles in each 
event with low $p_T$ selection leading to larger fluctuations. 

\begin{figure}
\vskip -2.5in
\epsfig{file=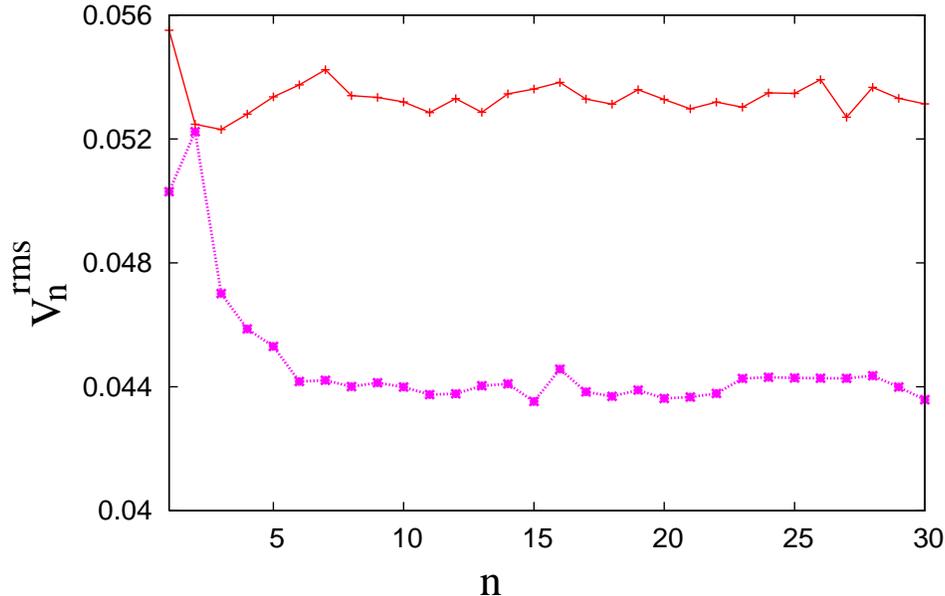,height=225mm}
\vskip -3.0in
\caption{These plots represent the root mean square values of the 
Fourier coefficients of the momentum anisotropy directly calculated using 
momenta of final particles from HIJING (with 15000 events). Dotted plot 
corresponds to particles with no momentum cuts, while the solid plot 
corresponds to the case when only particles with low $p_T (\le $ 200 MeV) 
are included.}
\label{fig:fig4}
\end{figure}

 In Fig.5 we replot the values of $v_n^{rms}$ (corresponding to the solid 
plot in Fig.3) with the inclusion of various physical factors such as finite
acoustic horizon and coherence of fluctuations as discussed in previous
sections. The solid curve shows plot of values of 
$v_n^{rms}$ (as given by the solid curve in Fig.3) with the inclusion of 
suppression factor (Eq.(5)) for superhorizon fluctuations, and the  dashed 
curve includes this suppression factor as well as the $cos\omega \tau$ 
oscillatory factor for sub-horizon anisotropies. The dotted curve, which is 
close to zero, shows the plot of average values of $v_n$ (instead of root 
mean square value) corresponding to the solid plot in Fig.3. 
Note that various plots here represent smooth joining of the points
obtained at integer $n$. This is why the dashed plot in Fig.5 which
includes $cos \omega \tau$ factor does not reach zero (which, from
Eq.(6) will typically occur at fractional values of $n$).

\begin{figure}
\vskip -2.5in
\epsfig{file=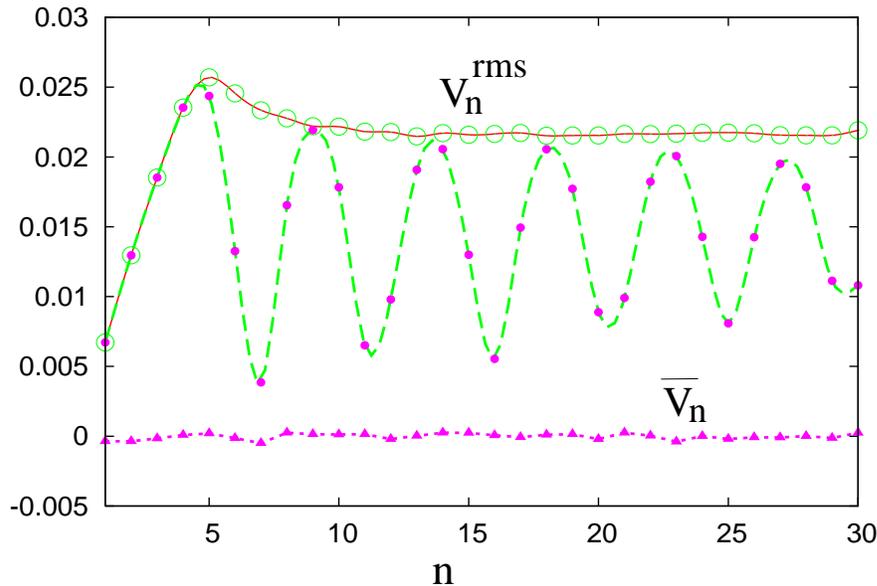,height=225mm}
\vskip -3.0in
\caption{These plots are obtained from the values of $v_n^{rms}$ of the 
solid plot in Fig.3  by including various factors representing the physics 
of acoustic horizon and coherence etc. as discussed in the text. Solid 
curve shows the plot which includes the suppression factor given in 
Eq.(5), while the dashed curve includes an additional $cos \omega \tau$ 
factor. Dotted curve represents the average values $\bar v_n$ 
(corresponding to solid plot in Fig.3). Note that the plots in this figure
only show the modeling of the type of suppression  factor and oscillations 
discussed in the text, which are superimposed on the solid plot in Fig.3.}
\label{fig:fig5}
\end{figure}

   The plots in Figs3-5 have several important features. First note, in 
Fig.3, a systematic decrease in the values of $v_n^{rms}$ as $n$ increases, 
starting from a significantly non-zero value. The errors in the values
of $v_n^{rms}$ are very small (e.g. less than about 2\% for the solid
plot in Fig.3). The overall shape of this plot may 
contain non-trivial information about the early stages of the system and 
its evolution. For example, the flattening of the curve by $n \simeq 10$
may be indicative of a qualitative change in the distribution of the 
corresponding initial anisotropies with wavelengths smaller than about 2 
fm (i.e. with wavelengths, as given after Eq.(5) with ${\bar R}^{fr}$ 
replaced with ${\bar R}(\tau_{eq})$ = 3 fm as the plots in Fig.3 are 
obtained from $F_n$s at $\tau_{eq}$). As we discussed above, the flat shape
of the dotted curve in Fig.3 corresponding to uniformly distributed
partons, and relatively smaller values of $v_n^{rms}$ for this case,
imply that statistical fluctuations due to finite particle numbers may
not dominate over the non-trivial shape of the solid plot in Fig.3.
As we discussed above,  the non-trivial shape of the dotted plot in Fig.4
(for direct momentum anisotropy of the final particles from HIJING) is
due to high $p_T$ particles. This is evidenced by the flatter shape of the
solid curve in Fig.4 where only low $p_T$ particles are included. The 
situation for the solid plot in Fig.3 (and hence for the plots in Fig.5) 
is different. Here the non-trivial shape of the plot does not change much
even when $p_T$ values of initial partons is constrained to be low
(e.g. $\le $ 350 MeV, lowering this value further leaves very few partons 
as initial partons do not have very low $p_T$ for center of mass energies
used here.). Such a non-trivial shape of the solid curve in Fig.3 should, 
therefore, directly relate to non-trivial nature of fluctuations present 
initially.

Another important aspect, which may also be reasonably robust in our model, 
is the first peak shown by the solid plot in Fig.5 at $n \sim 5$. 
As we had seen, the suppression factor in Eq.(5) becomes one for $\lambda 
\le \lambda_{max}$, with $\lambda_{max}$ corresponding to $n \simeq 5$ 
mode. Note that the exact location of this peak is determined 
by the value of $n_{min}$ (Eq.(6)) as well as the shape of the 
solid curve in Fig.3. This peak contains 
information about the freezeout stage, being directly related to the sound 
horizon size at that stage. (In this sense, this peak is similar to the 
first peak observed for CMBR anisotropies, which contains information about
the stage of decoupling \cite{cmbr}.) The dashed plot denoting oscillatory 
behavior will be the thing to look for and the plot (which only models the 
oscillations) shows that with enough statistics, one should be able to 
observe it if it is present. We emphasize again that our assumption of 
using the same proportionality constant (0.2) for all $v_n$s only affects 
the overall shape of the solid curve in Fig.3, and consequently the shapes 
of other two curves in Fig.5. It has no bearing on the existence of any of 
the peaks, especially the first peak. Average values ${\overline v_n}$ of 
$v_n$, shown by the dotted curve in Fig.5, are close to zero as expected. 
This also shows that we have enough statistics to control fluctuations 
due to statistical errors. We have not discussed any dissipative effects
which are known to affect the transverse expansion, hence the flow, in
crucial manner \cite{dsptv}. For example, it is known that viscosity 
reduces elliptic flow \cite{vscs}. 
Presumably such effects could lead to decreasing amplitude of $v_n^{rms}$ 
for these plots for large $n$, especially one should expect damped
oscillations, hence decreasing heights for the peaks at large $n$.

For near central collisions (with impact parameter $b \simeq 2$ fm) the 
elliptic flow has been estimated earlier in simulations \cite{v2cntrl}. The 
standard deviation for values of $v_2$ is found to be 
$\sigma \simeq 0.013$ \cite{v2cntrl}. This is consistent with the 
experimental results \cite{cntrlex}. The plots of $v_n^{rms}$ in Fig.5 show 
that the value we obtain for $n = 2$ is in reasonably good agreement with 
these values quoted in the literature. Note that the agreement is better for 
curves in Fig.5 which include the suppression factor as given in Eq.(5), as 
compared to the solid curve in Fig.3 which does 
not include this factor. Note that we do not attempt to compare $\bar v_2$ 
since in ref.\cite{v2cntrl,cntrlex} $\bar v_2$  is obtained in the usual 
manner by  determining the event plane for events with small, but non-zero 
impact parameter. In contrast, in our case the impact parameter is strictly 
zero and no special directions are determined on event by event basis. 
Presumably the comparison of the standard deviations is not affected much 
by this difference.

  We have repeated these calculations by taking other measure of spatial 
anisotropies, for example given by a uniform fluid density approximation
and taking spatial extent in a given $\phi$ bin to be proportional to
the transverse energy in that bin. For this case, the overall numerical 
values increase by about 20\% while the overall shapes of the curves 
remain essentially similar. Therefore, the qualitative aspects of our 
results seem to be reasonably independent of the specific technique 
chosen to represent the spatial anisotropies. The effects of adding a 
smooth background for the transverse energy density \cite{hijing}, as
discussed in Sect.II, are as expected, it reduces fluctuations, leading to
an overall decrease in the values of $v_n^{rms}$. Decrease is stronger
at smaller $n$. However, the overall shape of the curves remains 
essentially unchanged. Thus the main features of our results
remain unaffected.  Also, since we calculate anisotropies in the energy
density weighted average of radial coordinate, only relative fraction of
this soft energy density to $\epsilon_{tr}$ (Eq.(1)) is relevant. As the 
fraction of this soft background is increased, the values of $v_n^{rms}$ 
further decrease. When the contribution of smooth background is of the 
same order as that given by Eq.(1) (which happens with $\rho_c \simeq$ 
3 GeV/fm$^2$) then values of $v_n^{rms}$ are reduced by about a factor of
4 for small $n$ and by a factor of about 3 for large $n$.

  We have checked the effects of varying various parameters on the shape
of these curves. We briefly quote some results here, a more detailed
investigation and analysis will be done in a future work. We have considered
a larger value of $\tau_{fr}$ (= 12 fm) (for a given ${\bar R}(\tau_{eq})$) 
and our results remain almost unchanged. Much larger values of $\tau_{fr}$ 
(= 20 fm) shifts the first peak to larger $n$ (with corresponding shifts
in the other peaks) due to increase in the
average transverse velocity, and the peak becomes less prominent primarily 
due to the flattening of the overall plot of $v_n^{rms}$ for large $n$.
Interestingly, the effects of changing $\tau_{eq}$ are more prominent in
our model. With $\tau_{eq} = 0.2$ fm, we get an increase in the values
of $v_n^{rms}$ for small $n$ and a decrease for large $n$ (compared to
the case with $\tau_{eq} = 1$ fm). An increase
in $v_n^{rms}$ for smaller $\tau_{eq}$ is natural to expect as fluctuations 
in parton positions are more localized with less free-streaming (Eq.(1)).
The flattening of the overall plot (as in Fig.3) happens at a larger $n$ 
for smaller $\tau_{eq}$. This possibly again corresponds to the changeover
in the behavior of fluctuations, now happening at much shorter wavelengths
(hence large $n$) due to less free streaming. This may also be responsible
for the smaller values of $v_n^{rms}$  for large $n$, as for such short
wavelengths, the fluctuations may be decreased due to the corresponding 
parton density being larger when free-streaming is less. Increasing 
$\tau_{eq}$ to 2.0 fm leads to an overall decrease in the values of
$v_n^{rms}$ (decrease is stronger at smaller $n$, about 20\%). Also, the
flattening of the curve now happens at smaller $n$. 

The peak position, as governed by $n_{min}$ given in Eq.(6) (combined with 
the shape of the decreasing overall curve of $v_n^{rms}$), is sensitive to 
the value of sound speed $c_s$. Decreasing $c_s$ shifts the first peak to 
larger $n$, with corresponding shifts in the successive peaks. Thus,
the location of these peaks can give important information about the
equation of state during the early stages.
Increasing rapidity window decreases the values of $v_n^{rms}$ for all $n$,
presumably due to larger number of particles leading to less fluctuations. 
(With $\Delta y = 2$ in our model, values of $v_n^{rms}$ decrease by about 
30 \%). Increasing center of mass energy also leads to an overall
decrease in the values of $v_n^{rms}$, again possibly due to larger number
of partons and hence reduced initial anisotropies. It also leads to a
slight shift of the peaks to larger $n$. This happens due to larger value 
of ${\bar R}$ as can be seen from Eq.(6). For a nucleus of smaller size,
($Cu$), one gets smaller number of partons
increasing the values of $v_n^{rms}$. It also slightly shifts the 
first peak to smaller $n$, due to smaller value of ${\bar R}$.
We should mention that these patterns are obtained with smaller statistics
and hence are crude.  One needs a more careful and detailed analysis of 
the effects of changing various parameters. For non-central collisions 
the plot of $v_n^{rms}$ (as given by the solid plot in Fig.3) shows a peak 
at $n = 2$ as expected. When combined with the suppression factor of Eq.(5), 
it results in a double peak structure for the solid curve of Fig.5, with 
similar effect  for the dashed curve of Fig.5.

 We emphasize here that we are proposing a simple method for the 
calculation of $v_n^{rms}$. These are obtained by direct calculation
of variances of the distributions of $v_n$ in the laboratory fixed frame.
In the context of elliptic flow, there have been several studies 
on extracting elliptic flow coefficient from two-particle azimuthal 
correlations, and how to separate the non-flow contributions 
\cite{nonflow}. Two particle azimuthal correlations, which are 
experimentally measured, contain contributions from non-flow effects such 
as jets, resonance decays, HBT correlations, final state interactions etc. 
Various methods have been discussed to separate out the non-flow
contributions to the azimuthal correlations\cite{nonflow}, e.g. 
using cumulant expansion of multiparticle azimuthal correlations \cite{cuml}. 
Our estimates of $v_n^{rms}$, (either by direct calculation of variance, 
or using two particle correlations), will contain such non-flow 
contributions (as, for example, discussed above
for the plots in Fig.4). Though, here as one is making
a plot $v_n^{rms}$ for a whole range of values of $n$, different non-flow 
contributions may affect different parts of this plot. A detailed
investigation of these issues is needed to separate out, or at least 
estimate the effects of these non-flow contributions. We hope to address
these issues in a future work.

\section{Conclusions}

 Important aspects of our model are that we argue that important information
about initial anisotropies of the system and their evolution in relativistic
heavy-ion collisions can be obtained by plotting the root mean square values 
of the Fourier coefficients $v_n^{rms}$ of the anisotropies in the 
fluctuations $\delta p/p$ of the particle momenta, calculated in a fixed
laboratory frame, starting from $n = 1$ upto large values of $n \simeq 30$. 
Note that $n = 30$ almost corresponds to wavelength of fluctuation $\lambda$ 
at the surface of the region, at $\tau_{fr}$, being of order 1 fm. (At 
$\tau_{eq}$ it will correspond to $\lambda \simeq$ 0.7 fm). Fluctuations
with wavelengths smaller than 1 fm cannot be treated within hydrodynamical 
framework, so we restrict attention within this range of $n$. (It may be
useful to plot $v_n^{rms}$ for a much larger range of $n$.
One will expect that beyond a critical value of $n$ the nature of the
curve should change in some qualitative manner indicating breakdown
of underlying hydrodynamical description for smaller modes. The wavelength
corresponding to that critical value of $n$ will determine the smallest
scale below which hydrodynamical description is not valid.) Important thing 
to note is that by taking very large number of (central) events, the root 
mean square values of $v_n$s can show any possible systematic variation,
for example initial rapid decrease of $v_n^{rms}$ with $n$ which
eventually flattens out for large $n$. One needs to develop a proper 
understanding of this behavior, (which presumably suggests a qualitative
change in the nature of the fluctuations for small wavelengths)  and of 
various factors affecting this. Further, there is the possibility of a peak 
near $n \sim 5$ (for $Au-Au$ collision at 200 GeV) and of subsequent peaks 
for larger $n$. If any of these non-trivial features are detected in the 
particle momentum spectra then it can open up a new way of accessing the 
information about initial stages of the matter produced. Just as for CMBR 
where the location of the first peak refers to the decoupling stage, here 
also the location of the first peak will give information about the 
freezeout stage, including the all important equation of state 
which could  distinguish a QGP phase from a hadronic phase.
For CMBR, successive peaks yield important information about baryon
content etc. which couple to the radiation and hence contribute to the 
acoustic oscillations. In the same way for RHICE, if successive peaks are
present, then they may give information about the detailed properties of
the matter present at that stage, and any dissipative factors etc. 
One important factor which can affect the shapes of these curves, especially
the peaks, is the nature and presence of the quark-hadron transition.
Clearly the duration  of any mixed phase directly affects the freezeout
time and hence the location of the first peak. More importantly, any
softening of the equation of state near the transition may affect locations
of any successive peaks and their relative heights. Effects of mass ordering
may also be important in this respect, especially for the behavior of
these plots at large $n$ corresponding to small $\lambda$ probing very
small scale fluctuations of the order of hadronic scale. 

If one does see 
even the first peak for RHICE then one very important issue relevant for 
CMBR can be studied with controlled experiments. It is the issue of horizon 
entering. For example, by changing the nuclear size and/or collision energy, 
one can arrange the situation when first peak occurs at different values of 
$n$. As we have emphasized above, even the overall shape of the curve (as 
given by the solid plot in Fig.3) will contain valuable information about 
the presence and evolution of the initial state anisotropies of the system 
resulting from ultra-relativistic heavy-ion collisions. It will be 
interesting if one can find a way to analyze the effects of such 
fluctuations along the longitudinal direction, i.e. with rapidity, by taking 
fixed azimuthal angle window. Due to rapid longitudinal expansion, 
presumably one will need to confine to a narrow window of rapidity.  As we 
discussed earlier, longitudinal fluctuations are
not expected to be superhorizon. Hence one may expect qualitatively 
different behavior in the plots of $v_n^{rms}$ for this case compared to the 
transverse case (this is apart from the other important differences due to 
very different dynamics of longitudinal and transverse expansions).
However, the situation may be different in the context of color glass
condensate models \cite{cgc}. In such models, initially a color field
is established between receding nuclei, which subsequently decays
in particles. If certain fluctuations are imprinted on such a color
field in the longitudinal direction, then even if those are causal
for the color field, for the resulting parton system, they may constitute
fluctuations of super-horizon wavelengths (as the decay of the color
field will involve different time scale than the time scale of establishing
the initial color field). This issue needs to be investigated in detail
as it may provide a distinctive signal 
for color glass condensate initial conditions.
All the considerations discussed in this paper can 
be straightforwardly applied to non-central events, as well as for 
collision of nuclei which are deformed. The only difference is that for 
these cases certain Fourier coefficients (with small values of $n$) will 
be most dominant, as discussed above. As for the case of elliptic flow, 
here also it is important to estimate and measure higher moments of
the flow coefficients to have a detailed understanding of the nature
of the original fluctuations. Recall that these play crucial role for
the CMBR analysis where non-Gaussian effects are very tightly constrained
in inflationary models of density fluctuations. Similarly, in RHICE also, 
the physical origin of original fluctuations will constrain and determine
various moments of the flow coefficients. Other issues for CMBR physics
such as the effects of the thickness of the surface of last scattering 
on CMBR anisotropy power spectrum will have obvious implications for
RHICE where also the width of the freezeout surface will affect the
plots of values of $v_n^{rms}$.

\section*{Acknowledgments}

  We are very grateful to R. Rangarajan for very useful comments and 
discussion. We also thank L. Sriramkumar, S. Digal, R. Ray, B. Layek,
V. Tiwari, and U. Gupta for useful comments.


\end{document}